# A Study for a Tracking Trigger at First Level for CMS at SLHC

C. Foudas, A. Rose, J. Jones and G. Hall

Imperial College London, Blackett Lab, SW7 2BW, London, UK.
c.foudas@imperial.ac.uk

*Abstract*

It is expected that the LHC accelerator and experiments will undergo a luminosity upgrade which will commence after several years of running. This part of the LHC operations is referred to as Super-LHC (SLHC) and is expected to provide beams of an order of magnitude larger luminosity ($10^{35} cm^{-2} sec^{-1}$) than the current design. Preliminary results are presented from a feasibility study for a First Level Tracking Trigger for CMS at the SLHC using the data of the inner tracking detector. As a model for these studies the current CMS pixel detector with the same pixel size and radial distances from the beam has been used. Monte Carlo studies have been performed using the full CMS simulation package (OSCAR) and the occupancy of such a detector at SLHC beam conditions has been calculated. The design of an electron trigger which uses both the calorimeter energy depositions and the pixel data to identify isolated electrons and photons has been investigated. Results on the tracker occupancy and the electron trigger performance are presented.

## I. INTRODUCTION

The Large Hadron Collider (LHC) at CERN is scheduled to commence operations in 2007 and provide the experiments with proton beams colliding every 25 nsec at a centre of mass energy of 14 TeV and a luminosity which will gradually increase to $10^{34}$ $cm^{-2}$ $sec^{-1}$. The CMS detector at the LHC is a general purpose collider detector. The momentum of charged particles is measured using a silicon tracker detector in a 4T magnetic field and their energies using electromagnetic and hadronic calorimeters. A muon system surrounds the detector.

Triggering at LHC imposes severe challenges on the experimenters [1]. Using the inelastic proton-proton cross section at 14 TeV (70 mb) one can estimate that on average 22 interactions occur every 25nsec when protons collide at the LHC. These events are superimposed on any exotic physics events. Using the theoretically expected Higgs production cross sections one can calculate that approximately only one in approximately ten thousand billion events will be interesting. This requires a sophisticated system to select events that have a discovery potential online. This is achieved by the CMS Trigger system comprising two components: The First Level Trigger (FLT) and the Higher Level Trigger system (HLT).

The FLT is a custom made fast processor system that employs pattern recognition and fast summing techniques to perform the first level of event selection without introducing dead-time. The FLT uses only the CMS calorimeter data (energy depositions) and muon data (stubs and tracks) as input for its algorithms and produces a yes/no decision 128 beam (FLT pipeline latency) crossings after the interaction (3.2 μsec). The output rate of the first level trigger is expected to be approximately 100 kHz. Triggering at first level is the most challenging part of the online data selection since it requires very fast electronics with a significant part placed inside the CMS detector. This introduces additional constraints for the FLT electronics. They have to limit the amount of data transferred out of the detector, be radiation hard and have very limited power consumption. This is the reason why a tracking trigger at first level is so challenging and also explains why tracking information is not included in the FLT at the present design of CMS. The HLT is a large computer farm based on fast commercial processors that use both the calorimeter and the tracker data to form their decision.

## II. CMS AT THE SUPER LHC

Detailed studies of signals possibly discovered at the LHC as well as searching for new signatures of exotic physics requires proton beams that are far more intense than those at LHC. Several upgrade scenarios of the LHC machine have been considered [2,3]. According to the most financially realistic scenario, the LHC will be upgraded to provide proton beams with luminosity up to $10^{35}$ $cm^{-2}$ $sec^{-1}$ colliding at twice the frequency (80 MHz) of the present design but with the same centre of mass energy. This machine design is commonly referred as the Super LHC and it is expected to be operational some time after 2015. A consequence of this design is that the particle flux through the detector will increase by a factor of 10 whist the backgrounds due to minimum bias events will increase by at least a factor of 5 (at 80 MHz). Hence, at SLHC luminosity, there will be about 100 proton-proton collisions occurring every 12.5 ns, in contrast to only 22 collisions for normal LHC operations. This imposes several requirements on the CMS detector:

  I. A complete redesign of the CMS silicon tracker due to radiation hardness, high occupancy and readout clock speed issues.
  II. It requires a new FLT that is compatible with an 80 MHz clock and is also capable of reporting the position of the objects found with resolution equal to $\Delta\eta \times \Delta\varphi = 0.087 \times 0.087$ in order to cope with the increased background level. It is generally accepted that the FLT output rate should be maintained at the current design level of 100 kHz in order to avoid a complete redesign of the upstream data acquisition system.

III. The total FLT pipeline must be shorter than 512 crossings (6.4μs for 80 MHz clock) determined by the length of the on-detector electromagnetic calorimeter digital pipeline, which should not change.

IV. The increased luminosity will also require a drastic revision of the existing trigger strategies. As an example, the performance of the muon based Level-1 Triggers suffers from poor transverse momentum resolution at high $p_T$, mainly due to multiple Coulomb scattering in the steel between chamber stations and limited lever arm used in the muon system. Hence, due to the shape of the underlying background distributions, the required FLT rate reduction cannot be achieved by increasing the thresholds even if one is willing to pay the price of cutting useful physics. A gain in background reduction can only come from improving the FLT algorithms for selection of physics signals. Furthermore at the SLHC the FLT efficiency and purity to select Higgs and other exotic signals will degrade due to the increase in the number of minimum bias events by at least a factor of 5. Given that the current FLT design already uses information from all detectors but the silicon tracker, any improvements in background reduction efficiency and purity can only come by including tracking information at the First Level Trigger. Hence, the design of a First Level Tracking trigger is one of the main CMS design requirements for running at SLHC.

## A. A rough estimate of the Tracker Occupancy at SLHC beam conditions

The rough estimate of the occupancy of a tracking detector at the SLHC has been obtained using the Pythia Monte Carlo program. The simulation included effects due to the 4 Tesla CMS magnetic field but ignored all other detector effects such as inactive material or charge sharing. The results of this study for 88 minimum bias events per crossing are shown in Table 1. The occupancy has been estimated for radii 8, 11, 14 cm and for detector sizes of $1cm^2$ and $1.28cm \times 1.28$ cm. As seen in Table 1 the occupancy varies between an average of 1.0 hits/($cm^2 \times 12.5nsec$) in the outer layer and 2.4 hits/($cm^2 \times 12.5nsec$) in the most innermost layer. These results have been cross-checked against those from previous CMS studies [4] which are shown in Fig. 1 versus radius, at LHC, for different transverse momentum cuts and magnetic field. Using Fig. 1 the occupancy is estimated to be 3-4 hits per 12ns per $1.28cm \times 1.28cm$ at radius 10 cm. This number is larger but consistent with our results. For a more precise estimate, the results of Table 1 should be increased by 20% since the average number of minimum bias events per crossing at the SLHC is expected to be 110.

Table 1: Tracker occupancy at SLHC for radius 8, 11 and 14 cm from the beam axis. The occupancy is also given per $1.28 \times 1.28 cm^2$ to compare with previous estimates.

| Radius (cm) | Number of hits per $1cm^2$ per 12.5nsec | Number of hits per $1.28 \times 1.28 cm^2$ per 12.5nsec |
|---|---|---|
| 8 | 2.4 | 3.9 |
| 11 | 1.5 | 2.4 |
| 14 | 1.0 | 1.6 |

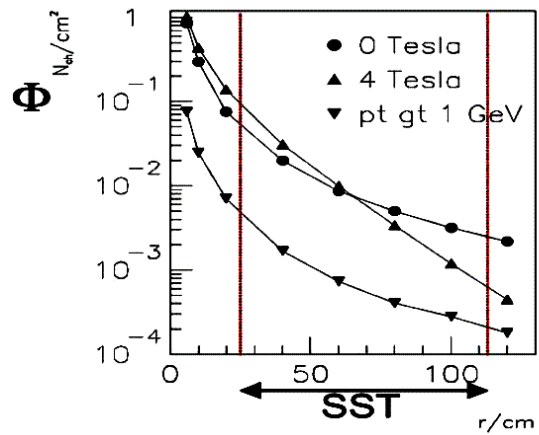

Figure 1: The occupancy (hits/$cm^2$/25nsec) versus radius from the beam axis computed using the CMS Monte Carlo for different values of the magnetic field and with and without a transverse momentum cut. These results have been computed for a luminosity of $10^{34}$ $cm^{-2}$ $sec^{-1}$. For SLHC these results need to be multiplied by a factor of 5.

## B. Expected Background Rejection from the First Level Tracking Trigger at the SLHC

Estimates of the level of background rejection achieved when one includes tracking information in the FLT can be obtained from studies of the HLT algorithms which use the tracker data in coincidence with calorimeter and muon detector data to select events [5]:

I. The isolated electron trigger rate can be suppressed by a factor of 10 by putting in coincidence electron candidates found in the calorimeter with hits found in the vertex detector. A further factor of 3 is gained by also including track stubs close to the calorimeter resulting in a factor of 30 in background rejection.

II. The τ-trigger suppresses backgrounds by a factor 10 when tracks in the vertex detector are put in coincidence with jets found in the calorimeter.

III. Muon triggers also benefit by a factor of 50 when the inner tracker information is put in coincidence with the outer First Level Muon candidates.

The conclusion from all these studies is that:
  I. Using tracking information in the FLT algorithms will more than compensate for the increase of the backgrounds.
  II. The largest benefit is obtained by using the tracks/hits and vertices found by the vertex detector.
  III. Prior information from other trigger systems such as the calorimeter or the muon system is vital to reduce the number of hit combinations in the inner tracker. This feature will have to be included in any algorithms implemented in a first level tracking trigger.

III. DESIGN CONSIDERATIONS FOR A FIRST LEVEL TRACKING TRIGGER AT THE SLHC

Based on the HLT studies shown before, a tracking trigger system is needed that provides for each crossing:

- Track-stubs and preliminary vertices from the vertex detector data.
- Track-stubs based on the outer tracker data.
- The Trigger should be free of dead-time. Hence, it has to process data every 12.5ns in a pipeline manner.
- To reduce the output data rate electrons, muons, taus, and jets found using the calorimeter and muon triggers must be correlated with track stubs and vertices found by the tracking trigger. An illustration of this is shown in Fig. 2.

IV. AN ELECTRON TRIGGER AT SLHC

A. The Electron finding algorithm

In references [5,6] an electron finding algorithm has been developed for the CMS 2.5 Level HLT and has been shown to be capable of significantly reducing the FLT isolated electron rate. The algorithm identifies electrons and rejects backgrounds from jets by associating hits in the pixel detector with electromagnetic clusters found in the calorimeter. It has been demonstrated to have an electron finding efficiency larger than 95% and reduces backgrounds from jets by a factor of 10. The logic of this algorithm initiates an electron search starting from an electromagnetic (EMC) cluster found in the calorimeter and searches for a consistent hit in the inner pixel layer. The hit search is restricted within a fixed window centred at the particle trajectory which starts from the nominal interaction point and ends at the centre of the electromagnetic cluster. The calculation of the particle trajectory takes in to account the CMS 4T magnetic field and the electromagnetic cluster energy. If no hits are found the event is rejected, otherwise a new trajectory and a z-vertex are calculated using the hit found and the EMC cluster coordinates. The new trajectory is used to search for hits in the next two pixel layers within tighter windows than before. If at least one of the pixel layers has a hit which is consistent with the new trajectory then the electromagnetic cluster is declared to be an electron otherwise it is rejected.

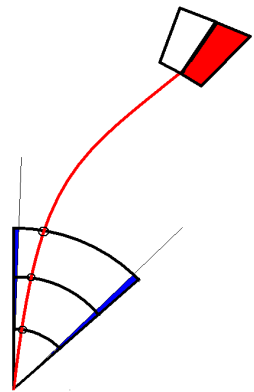

Figure 2: Hits in the vertex detector are correlated with electrons found by the calorimeter trigger.

The aim of this study was to investigate if such an electron finding algorithm can be simplified enough to be implemented in an ASIC or FPGA and therefore be used as part of the FLT decision at SLHC. The tools used for this study were the CMS Monte Carlo simulation and reconstruction Programs OSCAR and ORCA which include the full CMS detector simulation. The study was performed using a sample of 10000 Higgs events which decay into 4 electrons (H→ZZ→eeee). Minimum-bias events, corresponding to a luminosity of $2\times10^{33}$cm$^{-2}$sec$^{-1}$, have been superimposed on the Higgs events.

In the first part of this study the search window size and the occupancy are calculated using the electron trajectories from the Higgs decays. In the second part the window sizes and occupancies have been used to compute the number of hits per search window and estimate the number of combinations involved in reconstructing an electron-stub. In principle such electron-subs can be put in coincidence with isolated electrons found by the calorimeter trigger and produce a trigger which will have a higher rejection capability than the current isolated electron trigger which is only calorimeter based.

B. Search window sizes and Occupancy

In this study, following a similar logic to ref. [5,6], the coordinates of the most energetic crystal within the EMC cluster and the nominal interaction point are used to predict the trajectory of each of the four electrons. Hits[7] are then searched for in the 3 pixel layers in areas centred around the intersection points between the trajectories and the pixel surfaces. Shown in Fig. 3 are the hit distributions as a function of the absolute distance in r$\varphi$ of each hit from the intersection point of the trajectory with the pixel surface. The peak to the left side of the distributions represents hits associated with the electron track. The flat part of the distribution to the right of the peak represents hits unrelated to the electron tracks which are associated with backgrounds. As seen in Figure 3 the electron hit distributions fit well to a Gaussian distribution whilst the background is flat in r$\varphi$ as expected. Using Fig. 3, the search windows sizes in r$\varphi$ have been fixed to ±0.1, ± 0.15, ±0.20 cm for radii 4, 7 and 10 cm respectively. The z-width of the search window was chosen as ±15.00 cm based on the expected size of the proton bunch which is assumed to be the same for LHC and SLHC.

Using the flat part of the distributions of Fig. 3, which is due to backgrounds coming from minimum bias interactions, one can calculate more accurately the occupancy of the pixel detector at SLHC. The results of the calculation, which includes the magnetic field and all other detector effects are shown in Table 2 as a function of the pixel detector radius. The occupancy is given both for luminosity L=2×10$^{33}$/cm$^2$ sec at 40 MHz (fifth column) and for SLHC with L=10$^{35}$/cm$^2$ sec at 80 MHz (sixth column). The values shown here for SLHC are larger than those in Table 1 mainly due to the fact that this simulation includes detector effects which are not included in the calculation of Table 1. This calculation has also been repeated using a pure minimum bias Monte Carlo and gave identical results.

Table 2: The occupancy of the CMS pixel detector at LHC and SLHC for radii 4, 7 and 10 cm using the Higgs.

| R (cm) | Hits/ bin in plateau | Hits /bin /event | hits/bin /electron | Occup. (2×10$^{33}$) hits/cm$^2$/ 25 nsec | Occup. (1×10$^{35}$) hits/cm$^2$/ 12.5 nsec |
|---|---|---|---|---|---|
| 4 | 2500 | 0.250 | 0.0625 | 0.35 | 8.8 |
| 7 | 1100 | 0.110 | 0.0275 | 0.15 | 3.8 |
| 10 | 650 | 0.065 | 0.0162 | 0.09 | 2.3 |

### C. Window Occupancy and hit combinations

The Higgs Monte Carlo has also been used to compute the number of hits found within the search windows computed in the previous section. The hits within the electron peaks were integrated up to the window boundaries and the results are shown in Table 3. As shown in Table 3, for window areas of 6, 9 and 12 cm$^2$ corresponding to radii of 4, 7 and 10 cm, 3.1, 2.4 and 2.1 hits on average were found per window (Nhit/W). These hits include those from the Higgs decays. The occupancies from the previous section can be used to compute the number of background hits per window at L=2×10$^{33}$/cm$^2$ sec, given here as N33. This quantity scales with luminosity and at L=1×10$^{35}$ cm$^{-2}$sec$^{-1}$ is labelled N35. In Table 3, N33 and N35 are given as a function of the pixel detector radius.

As seen in Table 3 an enormous number of combinations is expected at SLHC and there is a need to reduce this number with on-detector zero-suppression techniques similar to those presented in [8].

This study has included electrons of energy larger than 3 GeV. However, in practice one is only interested in triggering electrons with transverse energy more than 30 GeV. Because these electrons bend less in the magnetic field, it is expected that the window sizes in r$\varphi$ will be inversely proportional to the electron transverse energy and hence the combinations should decrease accordingly.

Table 3: The search window occupancy as a function of the radius of the pixel layer for radii 4, 7 and 10 cm computed using 10000 Higgs→eeee events at a luminosity of L=2×10$^{33}$ cm$^{-2}$sec$^{-1}$. W- r$\varphi$, is the size of the windows in r$\varphi$ and W-A is the total window area. Nhit/W is the number of hits per window which includes the background hits as well as the hits from the Higgs decay electrons. N33 are the background hits only at L=2×10$^{33}$ cm$^{-2}$sec$^{-1}$ and N35 are the background hits at L=2×10$^{35}$ cm$^{-2}$sec$^{-1}$.

| R (cm) | W-r$\varphi$ (cm) | W-A cm$^2$ | Nhit / W | N33 | N35 |
|---|---|---|---|---|---|
| 4 | ± 0.10 | 6.0 | 3.1 | 2.1 | 53 |
| 7 | ± 0.15 | 9.0 | 2.4 | 1.4 | 34 |
| 10 | ± 0.20 | 12.0 | 2.1 | 1.1 | 27 |

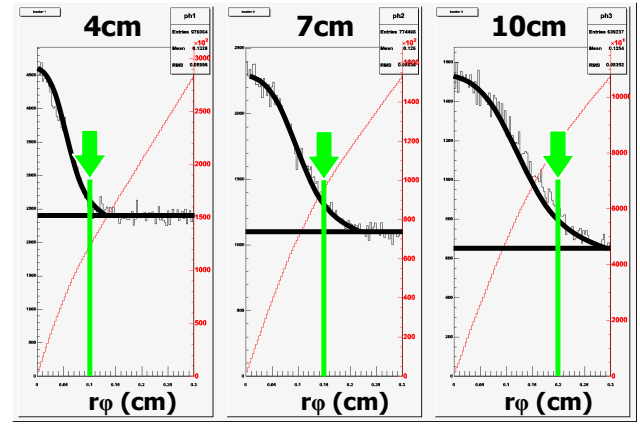

Figure 3: Hit distributions for the three CMS Pixel Layers at radii 4, 7, 10 cm. The histogram is the hit distribution and the red line is the integral of the hit distribution. Shown in green are the sizes of the 3 windows in r$\varphi$.

Shown in Fig. 4 is the fraction of the electrons from Higgs decays with a number of combinations, which is larger than the value shown on the horizontal axis. The combinations have been computed by including only the hits from the windows at r = 4, r = 7 cm. As seen here the large majority of the events (>70%) have more than 50 combinations.

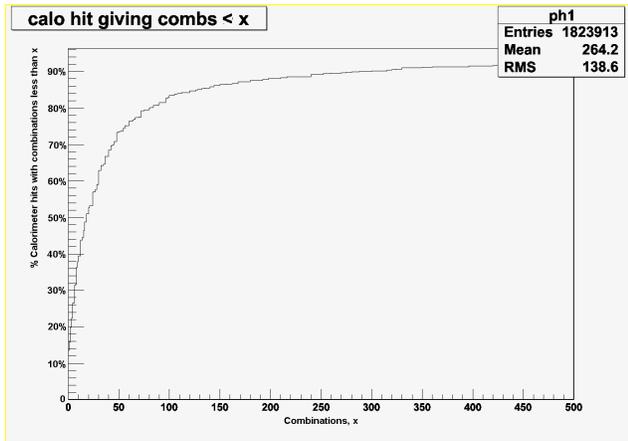

Figure 4: The percentage of electron tracks with number combinations larger than the value shown on the x-axis.

## V. CONCLUSIONS

A preliminary study has been carried out to investigate the challenge of building a tracking trigger at the first level for CMS at the SLHC.

The occupancy of a pixel detector at the SLHC has been estimated using several different techniques and varies between 8.8 hits/cm$^2$/12.5nsec at radius 4 cm and 2.3 hits/cm$^2$/12.5nsec at radius of 10 cm.

We have investigated the possibility of designing a tracking trigger that could produce electron-track stubs and put them in coincidence with electron objects found by the calorimeter trigger. In our model this was done by searching for hits within fixed windows centred at the electron trajectory which was fixed by the electromagnetic object found by the calorimeter trigger and the nominal interaction point. The number of hit combinations from the two outer pixel layers was estimated to be of the order of $10^3$. We conclude that the number of combinations needs to be reduced either by novel on-detector techniques such as those of ref. [8] or by increasing the electron transverse energy cuts.